\documentclass[11pt,a4paper]{article}
\pdfoutput=1

\usepackage[utf8]{inputenc} 
\usepackage{a4wide}
\usepackage{amsmath}
\usepackage{cite}
\usepackage{xcolor}
\usepackage{amssymb}
\usepackage{amsfonts}
\usepackage{graphicx}
\usepackage{hyperref}
\usepackage{mathbbol}
\usepackage{amssymb}
\usepackage{booktabs}
\usepackage{multirow}
\usepackage{appendix}

\DeclareSymbolFontAlphabet{\amsmathbb}{AMSb}

\begin{document}
\begin{titlepage}

${}$\vskip 3cm
\vspace*{-15mm}
\begin{flushright}
SISSA 25/2018/FISI\\
IPMU18-0121
\end{flushright}
\vspace*{0.7cm}

\vskip 9mm
\begin{center}
{\bf\Large 
Lepton Masses and Mixing from Modular~$\boldsymbol{S_4}$ Symmetry
} \\[8mm]
J.~T.~Penedo$^{~a}$,
and
S.~T.~Petcov$^{~a,b,}$\footnote{Also at:
Institute of Nuclear Research and Nuclear Energy,
Bulgarian Academy of Sciences, 1784 Sofia, Bulgaria.} \\
\vspace{8mm}
$^{a}$\,{\it SISSA/INFN, Via Bonomea 265, 34136 Trieste, Italy} \\
\vspace{2mm}
$^{b}$\,{\it Kavli IPMU (WPI), University of Tokyo, 5-1-5 Kashiwanoha, 277-8583 Kashiwa, Japan}
\end{center}
\vspace{8mm}

\begin{abstract}
We study models of lepton masses and mixing
based on broken modular invariance.
We consider invariance under the finite 
modular group $\Gamma_4 \simeq S_4$
and focus on the minimal scenario where
the expectation value of the modulus
is the only source of symmetry breaking,
such that no flavons need to be introduced.
After constructing a basis for the lowest weight modular forms,
we build two minimal models,
one of which successfully accommodates 
charged lepton masses and neutrino oscillation data,
while predicting the values of the Dirac and Majorana CPV phases.
\end{abstract}

\end{titlepage}

\setcounter{footnote}{0}
\setcounter{page}{2}

\section{Introduction}
\label{sec:intro}

It is a challenge to understand the origin of the 
flavour structure of quarks and leptons.
Data obtained in neutrino oscillation experiments 
(see, e.g.,~\cite{Patrignani:2016xqp})
has revealed a pattern of large mixing in the lepton sector, 
which is being probed with increasing precision.
Additionally, a preference for both a neutrino mass spectrum with 
normal ordering (NO)
and for a Dirac CP violation (CPV) phase close to maximal
have been found in recent global fits (see, e.g.,~\cite{Capozzi:2018ubv}).

A popular approach to explaining the observed mixing patterns
consists in assuming the presence of a spontaneously broken 
discrete flavour symmetry.
Typically, however, such breaking requires 
the introduction of a plethora of scalar fields (flavons)
with a set of particularly aligned vacuum expectation values (VEVs).
Arranging for such an alignment 
usually calls for the construction of 
rather ``baroque'' scalar potentials.

An interesting framework for the construction of flavour models
has recently been put forward in Ref.~\cite{Feruglio:2017spp}
and further explored in Ref.~\cite{Kobayashi:2018vbk}.
In this setup, invariance under finite subgroups of the 
modular group shapes the relevant mass matrices.
These models generalise existing approaches to the flavour puzzle,
which can be seen as the particular cases of setting modular weigths to zero.
In the most economical versions of these models, 
the VEV of the modulus $\tau$ is the only source of symmetry breaking,
bypassing the need for flavon fields. 
Leading-order predictions may be derived by treating $\tau$ as a spurion.

In the present paper we investigate the consequences of 
enforcing modular invariance in the lepton sector. 
We focus on the action of the finite modular 
group $\Gamma_4$, which is isomorphic to 
the group of permutations of four objects $S_4$.
After reviewing the necessary formalism, we explicitly construct the
generators of modular forms of level $N=4$. We then investigate 
two minimal models where neutrino masses arise from the Weinberg operator
and where no flavons are introduced.

\section{The Setup}
\label{sec:setup}

Following Ref.~\cite{Feruglio:2017spp}, we consider 
the infinite groups $\Gamma(N)$,
\begin{align}
\Gamma(N) \equiv \left\{ \gamma =
\begin{pmatrix}
a &b \\ 
c &d 
\end{pmatrix}
\,\middle|\,\, a,b,c,d \in \amsmathbb{Z}
\,\wedge\,\det \gamma = 1
\,\wedge\,\gamma = 
\begin{pmatrix}
1 &0 \\ 
0 &1 
\end{pmatrix}\, (\text{mod }N)
 \right\}\,,
\label{eq:gammaN}
\end{align}
where $N$ is a positive integer. The group 
$\Gamma(1) \simeq \text{SL}(2,\amsmathbb{Z})$ 
is the modular group and $\Gamma(N>1)$ are normal 
subgroups of $\Gamma(1)$. Taking the quotient of 
$\Gamma(1)$ and $\Gamma(2)$ by $\{\mathbb{1},-\mathbb{1}\}$ 
we obtain the groups of linear fractional transformations, 
$\overline\Gamma(N)\equiv \Gamma(N)/\{\mathbb{1},-\mathbb{1}\}$ 
for $N=1,2$, and $\overline\Gamma(N>2)\equiv \Gamma(N)$. 
Elements of $\overline \Gamma(N)$ act on a complex variable $\tau$ as:
\begin{align}
\tau \,\rightarrow\,
\gamma \tau = \frac{a\tau+b}{c\tau+d}
\,,\quad \text{with }\gamma =
\begin{pmatrix}
a &b \\ 
c &d 
\end{pmatrix}\in \,\overline \Gamma(N)\,,
\end{align}
%
and it can be shown that the upper half-plane
$\{\tau \in \amsmathbb{C}\,|\, \text{Im}(\tau) >0 \}$
is mapped to itself under this action.
The complex variable $\tau$ is henceforth restricted 
to have positive imaginary part.

We are interested in studying physical actions which are invariant 
under transformations of the finite modular groups $\Gamma_N$. 
These discrete groups are  obtained from
the quotient of two of the aforementioned infinite groups, 
namely $\Gamma_N \equiv \overline\Gamma(1)/\overline\Gamma(N)$. 
The group $\overline\Gamma(1)$ is generated by two elements $S$ and $T$
acting on $\tau$ as
\begin{align}
S:\,\,\tau \,&\rightarrow\,
-1/\tau
\,,\qquad S = \begin{pmatrix}
0 & 1 \\ 
-1 & 0
\end{pmatrix}\,,\\
T:\,\,\tau \,&\rightarrow\,
\tau +1
\,,\qquad T = \begin{pmatrix}
1 & 1 \\ 
0 & 1
\end{pmatrix}\,,
\end{align}
%
and satisfying
\begin{align}
S^2 \,=\, (ST)^3 \,=\, \mathbb{1}\,.
\end{align}
%
The presentation of the groups $\Gamma_N$ can be obtained from that 
of  $\overline\Gamma(1)$ by extending it with the condition 
$T^N = \mathbb{1}$, cf.~Eq.~\eqref{eq:gammaN}. 
Thus, the generators $S$ and $T$ of $\Gamma_N$ obey
\begin{align}
S^2 \,=\, (ST)^3 \,=\, T^N \,=\, \mathbb{1}\,.
\label{eq:presentN}
\end{align}
%
\vskip 2mm

We consider modular-invariant
$\mathcal{N}=1$ global supersymmetric (SUSY) 
actions~\cite{Ferrara:1989bc,Ferrara:1989qb},
\begin{align}
\mathcal{S} \,=\,
\int d^4x \,d^2\theta\,d^2\overline\theta \,\,K(\chi_i,\overline\chi_i;\tau,\overline\tau)
+\int d^4x \,d^2\theta \,\,W(\chi_i;\tau) + \text{h.c.}\,,
\end{align}
%
where $\chi_i$ denotes the set of matter chiral superfields of the theory
\footnote{As in Ref.~\cite{Feruglio:2017spp},
we turn off gauge interactions
and treat $\tau$ as a dimensionless spurion.
}.
The physical action $\mathcal{S}$ is required to be invariant
under the action of $\Gamma_N$. 
Under an element of this group,
one has the following transformations of $\tau$ and of the fields $\chi_i$:
\begin{align}
\begin{cases}
\,\tau \,\rightarrow\,
\gamma \tau = \dfrac{a\tau+b}{c\tau+d}
\\[5mm]
\,\chi_i \,\rightarrow\,\left(c \tau+d \right )^{-k_i}\rho_i(\gamma)\,\chi_i 
\end{cases},
\qquad \text{with }\gamma =
\begin{pmatrix}
a &b \\ 
c &d 
\end{pmatrix}\in \,\Gamma_N\,,
\label{eq:fields}
\end{align}
%
where $\rho_i$ are unitary representation matrices
and the $k_i$ are integers.
The fields $\chi_i$ are said to carry weigth $-k_i$.
We require that the superpotential $W$ remains invariant under $\Gamma_N$
and that the Kähler potential is changed at most by a Kähler transformation.
To satisfy this last condition, we work with the Kähler:
\begin{align}
K(\chi_i,\overline\chi_i;\tau,\overline\tau)= -h\, \Lambda_0^2\, \log (-i(\tau-\overline\tau)) + \sum_i\frac{|\chi_i|^2}{(-i(\tau-\overline\tau))^{k_i}}\,,
\end{align}
%
with $h>0$ and $\Lambda_0$ a mass parameter.
After $\tau$ develops a VEV, it gives rise to kinetic 
terms for the matter fields,
\begin{align}
\mathcal{L}
\,\supset\,
\sum_i\frac{\partial_\mu\overline\chi_i\,\partial^\mu\chi_i}{(2\,\text{Im}\langle\tau\rangle)^{k_i}}\,.
\end{align}
%
These terms can be made canonical by rescaling the fields $\chi_i$,
which in practice amounts to a redefinition of superpotential parameters. 
The superpotential reads
\begin{align}
W(\chi_i;\tau)= 
\sum_n
\sum_{\{i_1,\,\ldots,\,i_n\}}
\left(
Y_{\{i_1,\,\ldots,\,i_n\}}(\tau)\,\,\chi_{i_1}\ldots\chi_{i_n}
\right)_\mathbf{1}
\,,
\label{eq:W}
\end{align}
%
and should remain unchanged under $\Gamma_N$.

Given the field transformations in Eq.~\eqref{eq:fields}, 
the functions $Y_{\{i_1,\,\ldots,\,i_n\}}(\tau)$ in \eqref{eq:W}
must transform under $\Gamma_N$ as (we omit the indices $i_1,\,\ldots,\,i_n$):
\begin{align}
\begin{cases}
\,\tau \,\rightarrow\,
\gamma \tau = \dfrac{a\tau+b}{c\tau+d}
\\[5mm]
\,
Y(\tau)\,\rightarrow\,
Y(\gamma\tau) = 
\left(c \tau+d \right )^{2k_Y}\rho_Y(\gamma)\,Y(\tau)
\end{cases},
\qquad \text{with }\gamma =
\begin{pmatrix}
a &b \\ 
c &d 
\end{pmatrix}\in \,\Gamma_N\,,
\label{eq:modular}
\end{align}
%
where the unitary matrix $\rho_Y(\gamma)$ and the non-negative 
integer $k_Y$, as will shortly be shown, must satisfy specific conditions.

Holomorphic functions $f(\tau)$ satisfying 
$f(\gamma\tau) = (c\tau+d)^{2k} f(\tau)$
with $\gamma \in \Gamma(N)$ and $k \in \amsmathbb{N}_0$ are said 
to be modular forms of weight $2k$ and level $N$. For $k=0$, the modular 
forms are constant functions, while for $k<0$ modular forms do not exist.
Modular forms are important objects in the present construction
since, under $\Gamma_N$, modular forms of weight $2k_Y$ and level $N$
transform in the way we require $Y(\tau)$ in Eq.~\eqref{eq:modular} 
to transform.
The requirement of invariance of the superpotential then implies that
the functions $Y(\tau)$ are modular forms of level $N$. Their weights
must cancel those of the fields in Eq.~\eqref{eq:W},
$2k_Y-k_{i_1}-\ldots-k_{i_n}=0$. Additionally, the tensor product 
$\rho_{Y}\otimes\rho_{i_1}\otimes\ldots\otimes\rho_{i_1}$ should contain 
at least one singlet. The subscript $\mathbf{1}$ in Eq.~\eqref{eq:W} 
indicates a sum (with independent coefficients) of
all possible singlets one can extract from this product.

\vskip 2mm

The effects of enforcing invariance under $\Gamma_N$ for $N=2$ and $N=3$ 
have been studied in Refs.~\cite{Feruglio:2017spp,Kobayashi:2018vbk}. 
The group $\Gamma_2$ is isomorphic to $S_3$ while $\Gamma_3 \simeq A_4$. 
In what follows we consider invariance under the group 
$\Gamma_4 \simeq S_4$, whose generators satisfy the following
 presentation rules:
\begin{align}
S^2 \,=\, (ST)^3 \,=\, T^4 \,=\, \mathbb{1}\,.
\label{eq:present4}
\end{align}
%
For a generic value of $\langle\tau\rangle$, 
$\Gamma_4$ will be fully broken.
It is important to remark that the leading-order results obtained with 
this approach are susceptible to corrections from a small number of sources, 
namely from SUSY breaking and corrections to the Kähler potential. 
In generalisations of our approach where flavons are introduced, 
additional corrections are expected from vacuum (mis)alignment.

\subsection{Generators of Modular Forms of Level $\boldsymbol{N=4}$}
\label{sec:generators}

The functions $Y(\tau)$ are modular forms of level $N=4$ and weight $2k_Y$. 
The dimension of the space of modular forms of level 4 and weight $2k$
is $4k+1$. Thus, the space of (level 4) forms which carry 
the lowest nontrivial weight, $2k=2$, has dimension $4k+1 = 5$. 
It proves useful to explicitly find a 
basis $\{Y_1(\tau),\ldots,Y_5(\tau)\}$ of this lowest weight space,
since modular forms of higher weights can be constructed from
homogeneous polynomials in these five modular forms $Y_i$ ($i=1,\ldots,5$).

A starting point in this search is the recognition of certain 
properties of the Dedekind eta function 
$\eta(z \in \amsmathbb{C})$, defined as:
\begin{align}
\eta(z) \,\equiv\, q^{1/24}\,\prod_{k=1}^\infty\, (1-q^k)\,,\qquad\text{with }q= e^{2\pi i\, z}\,.
\label{eq:Deta}
\end{align}
The $\eta$ function satisfies 
$\eta(z+1) = e^{i\pi/12}\,\eta(z) $
and $\eta(-1/z)=\sqrt{-iz}\,\eta(z)$,
as well as the identity (see, e.g.,~\cite{Kohler:2011eta}):
\begin{align}
\eta\left(z+\frac{1}{2}\right) = 
e^{i\pi/24}\,
\frac{\eta^3(2z)}{\eta(z)\eta(4z)}
\,.
\label{eq:cubic}
\end{align}
%

One sees that the set
\begin{align}
\left\{\eta_i\right\} = 
\left\{
\eta\left(\tau+\frac{1}{2}\right),\,
\eta\left(4\tau\right),\,
\eta\left(\frac{\tau}{4}\right),\,
\eta\left(\frac{\tau+1}{4}\right),\,
\eta\left(\frac{\tau+2}{4}\right),\,
\eta\left(\frac{\tau+3}{4}\right)
\right\}
\label{eq:set}
\end{align}
%
respects a certain notion of closure under the action of
$\Gamma_4$ generators $S$ and $T$,
since under their action one has:
\begin{align}
S:\quad
\left\{\begin{array}{@{\quad}r@{\quad\rightarrow\quad}l@{}}
\eta\left(\tau+\dfrac{1}{2}\right) &
\dfrac{1}{\sqrt{2}}\,\sqrt{-i\tau}\,\eta\left(\dfrac{\tau+2}{4}\right) \\[4mm]
\eta\left(4\tau\right) &
\dfrac{1}{2}\,\sqrt{-i\tau}\,\eta\left(\dfrac{\tau}{4}\right) \\[4mm]
\eta\left(\dfrac{\tau}{4}\right) &
2\,\sqrt{-i\tau}\,\eta\left(4\tau\right) \\[4mm]
\eta\left(\dfrac{\tau+1}{4}\right) &
e^{-i\pi/6}\,\sqrt{-i\tau}\,\eta\left(\dfrac{\tau+3}{4}\right) \\[4mm]
\eta\left(\dfrac{\tau+2}{4}\right) &
\sqrt{2}\,\sqrt{-i\tau}\,\eta\left(\tau+\dfrac{1}{2}\right) \\[4mm]
\eta\left(\dfrac{\tau+3}{4}\right) &
e^{i\pi/6}\,\sqrt{-i\tau}\,\eta\left(\dfrac{\tau+1}{4}\right)
\end{array}\right.
\,,
\label{eq:etaS}
\end{align}
%
and
\begin{align}
T:\quad
\left\{\begin{array}{@{\quad}r@{\quad\rightarrow\quad}l@{}}
\eta\left(\tau+\dfrac{1}{2}\right) &
e^{i\pi/12}\,\eta\left(\tau+\dfrac{1}{2}\right) \\[4mm]
\eta\left(4\tau\right) &
e^{i\pi/3}\,\eta\left(4\tau\right) \\[2mm]
\eta\left(\dfrac{\tau}{4}\right) &
\eta\left(\dfrac{\tau+1}{4}\right) \\[4mm]
\eta\left(\dfrac{\tau+1}{4}\right) &
\eta\left(\dfrac{\tau+2}{4}\right) \\[4mm]
\eta\left(\dfrac{\tau+2}{4}\right) &
\eta\left(\dfrac{\tau+3}{4}\right) \\[4mm]
\eta\left(\dfrac{\tau+3}{4}\right) &
e^{i\pi/12}\,\eta\left(\dfrac{\tau}{4}\right)
\end{array}\right.
\,.
\label{eq:etaT}
\end{align}
%
The transformations under $S$ of the elements 
$\eta(\tau + 1/2)$ and $\eta((\tau +2)/4)$ can be derived by making 
use of the relation~\eqref{eq:cubic}.
Up to multiplicative factors, this set is closed under $S$ and $T$.
Furthermore, each element is taken into itself (up to a factor) by 
the (left-)actions of $S^2$, $(ST)^3$ and $T^4$.
The above suggests that the desired $k=1$ modular forms can be written
as linear combinations of the logarithmic derivatives of the elements
of the set $\left\{\eta_i\right\}$. We define:
\begin{align}
Y(a_1,\ldots,a_6|\tau) &\equiv \frac{d}{d\tau}\left(\sum_{i=1}^6 \,a_i\,\log \eta_i(\tau)\right)\nonumber \\
&= a_1 \frac{\eta'(\tau+1/2)}{\eta(\tau+1/2)}
 + 4\,a_2 \frac{\eta'(4\tau)}{\eta(4\tau)}
 + \frac{1}{4}\bigg[a_3 \frac{\eta'(\tau/4)}{\eta(\tau/4)}
 \\
&\, + a_4 \frac{\eta'((\tau+1)/4)}{\eta((\tau+1)/4)}
 + a_5 \frac{\eta'((\tau+2)/4)}{\eta((\tau+2)/4)}
 + a_6 \frac{\eta'((\tau+3)/4)}{\eta((\tau+3)/4)}\bigg].
 \nonumber
\end{align}
%
The use of logarithmic derivatives
allows one to eliminate the multiplicative factors in the 
transformations \eqref{eq:etaS} and \eqref{eq:etaT} 
by requiring $\sum_i a_i = 0$.
We are thus left with five independent linear combinations of the 
$\eta_i'/\eta_i$, as expected. Under the action of the generators 
$S$ and $T$ this function transforms as:
\begin{align}
S:\,\,
Y(a_1,\ldots,a_6|\tau)\,\,&\rightarrow\,\,
Y(a_1,a_2,a_3,a_4,a_5,a_6|{-1/\tau})\,=\,
\tau^2 \,
Y(a_5,a_3,a_2,a_6,a_1,a_4|\tau)
\,,
\\
T:\,\,
Y(a_1,\ldots,a_6|\tau)\,\,&\rightarrow\,\,
Y(a_1,a_2,a_3,a_4,a_5,a_6|\tau+1)\,=\, 
Y(a_1,a_2,a_6,a_3,a_4,a_5|\tau)
\,.
\end{align}
%

We find a basis for lowest weight modular forms,
\begin{align}
\begin{array}{@{}r@{\,\equiv\,}l@{}}
Y_1(\tau) & Y(1,1,\omega,\omega^2,\omega,\omega^2|\tau)\,,\\[2mm]
Y_2(\tau) & Y(1,1,\omega^2,\omega,\omega^2,\omega|\tau)\,,\\[2mm]
Y_3(\tau) & Y(1,{-1},{-1},{-1},1,1|\tau)\,,\\[2mm]
Y_4(\tau) & Y(1,-1,-\omega^2,-\omega,\omega^2,\omega|\tau)\,,\\[2mm]
Y_5(\tau) & Y(1,-1,-\omega,-\omega^2,\omega,\omega^2|\tau)\,,
\end{array}
\label{eq:Yi}
\end{align}
%
where $\omega \equiv e^{2\pi i/3}$. These five linearly independent 
forms $Y_i (\tau)$ arrange themselves into two irreducible 
representations of $\Gamma_4 \simeq S_4$,
a doublet $\mathbf{2}$ and a triplet $\mathbf{3'}$,
\begin{align}
Y_\mathbf{2}(\tau) \,\equiv\,
\begin{pmatrix}Y_1(\tau)\\ Y_2(\tau)\end{pmatrix}\,,
\qquad
Y_\mathbf{3'}(\tau) \,\equiv\,
\begin{pmatrix}Y_3(\tau)\\ Y_4(\tau)\\ Y_5(\tau)\end{pmatrix}\,.
\end{align}
%
The multiplets $Y_\mathbf{2}$ and $Y_\mathbf{3'}$ transform 
under $\Gamma_4$ as indicated in Eq.~\eqref{eq:modular}.
In Appendix~\ref{app:basis} we specify our basis choice
for the representation matrices $\rho(\gamma)$ of $S_4$
and we list the Clebsch-Gordan coefficients for this basis
in Appendix~\ref{app:cbc}.
In Appendix~\ref{app:qexp} we give the $q$-expansions of 
(combinations of) the five functions defined in Eq.~\eqref{eq:Yi}.

Multiplets transforming in the other representations of $S_4$
can be obtained from tensor products of $Y_\mathbf{2}$ and $Y_\mathbf{3'}$.
The representations $\mathbf{1}$ and $\mathbf{3}$ arise at weight $2k = 4$,
while the $\mathbf{1'}$ representation first arises at weight 6.
Since we can form 15 combinations $Y_i Y_j$,
one may expect 15 independent (level 4) forms at weight $2k=4$.
However, the dimension of the space of these forms is $4k+1=9$.
In fact, we find 6 constraints between the several $Y_i Y_j$, 
which we list in Appendix~\ref{app:constraints}.
These constraints reduce the 15 potentially independent combinations 
to 9 truly independent ones, which are organised in the following 
representations of $S_4$:
\begin{equation}
\begin{aligned}
Y^{(4)}_\mathbf{1}    &\,=\, Y_1 Y_2 \,\sim\, \mathbf{1}  \,,\\[2mm]
{Y^{(4)}_\mathbf{2}}  &\,=\, (Y_2^2,\, Y_1^2)^T \,\sim\, \mathbf{2}  \,,\\[2mm]
{Y^{(4)}_\mathbf{3}}  &\,=\, (Y_1Y_4-Y_2Y_5,\,Y_1Y_5-Y_2Y_3,\,Y_1Y_3-Y_2Y_4)^T \,\sim\, \mathbf{3}  \,,\\[2mm]
{Y^{(4)}_\mathbf{3'}}  &\,=\, (Y_1Y_4+Y_2Y_5,\,Y_1Y_5+Y_2Y_3,\,Y_1Y_3+Y_2Y_4)^T \,\sim\, \mathbf{3'}
\,.
\label{eq:9}
\end{aligned}
\end{equation}

\section{Phenomenology}
\label{sec:pheno}

To understand how invariance under the subgroup
$\Gamma_4$ of the modular group may play a role
in determining lepton masses and mixing,
one needs to specify the $S_4$ representations $\rho_i$
and the modular weights $-k_i$ of the relevant fields $\chi_i$,
which transform as indicated in Eq.~\eqref{eq:fields}.
In what follows, we search for choices of representations and weights
which are in line with a certain notion of minimality.

In a minimal approach, the superpotential $W$ includes only 
the Yukawa interactions of the Minimal Supersymmetric Standard Model (MSSM)
and the SUSY Weinberg operator as the origin of (Majorana) neutrino masses:
\begin{align}
W \,&=\,
\alpha \, \big(E^c\, L\, H_d\, f_E(Y_\mathbf{2},Y_\mathbf{3'}) \big)_\mathbf{1}
+ \frac{g}{\Lambda}\,\big(  L \,H_u\, L\, H_u\,
 f_W(Y_\mathbf{2},Y_\mathbf{3'})\big)_\mathbf{1}
 \,,
\label{eq:Wmin}
\end{align}
%
where, as mentioned before, different coefficients are 
implied for different singlet combinations.
No flavons are present in the above superpotential.
Recall also that the functions
$Y_\mathbf{2}=Y_\mathbf{2}(\tau)$ and $Y_\mathbf{3'} = Y_\mathbf{3'}(\tau)$
have definite transformation properties under $\Gamma_4$.
After the breaking of modular symmetry we obtain:
\begin{align}
W\, \rightarrow\,
\lambda_{ij} \, E_i^c\, (L_j\, H_d)
+ c_{ij}\,(L_i \,H_u)\,(L_j \,H_u)\,,
\end{align}
%
which leads to the Lagrangian terms
\begin{align}
\mathcal{L} \,\supset\,
-\frac{1}{2}\,\big(M_\nu\big)_{ij}\,\overline{\nu_{iL}}\,\nu_{jR}^c
-\big(M_e\big)_{ij}\,\overline{e_{iL}}\,e_{jR} + \text{h.c.}
\,,
\end{align}
%
written in terms of four-spinors,
where~\footnote{In the decoupling limit of the MSSM
(e.g.~when the heavier Higgs scalar states have masses exceeding $\sim 1$ TeV),
the lightest Higgs boson couplings to charged leptons
(and, for that matter, also to quarks)
differ insignificantly from those of the Standard Model Higgs.}
$M_e \equiv v_d\,\lambda^\dagger$ and $M_\nu \equiv 2\, c\, v_u^2$,
with $\langle H_u\rangle = (0,v_u)^T$ and $\langle H_d\rangle = (v_d,0)^T$.

The generic assignments of representations and weights to the 
MSSM fields present in Eq.~\eqref{eq:Wmin} are defined 
in Table~\ref{tab:assignments}.
\begin{table}[t]
\centering
\renewcommand{\arraystretch}{1.2}
\begin{tabular}{clccccc}
\toprule
  & & $H_u$ & $H_d$ & $L$ & $E^c_{1,2,3}$ \\ 
\midrule
\multicolumn{2}{c}{$\quad\text{SU}(2)_L \times \text{U}(1)_Y\quad$ }
 & $(\mathbf{2},1/2)$ & $(\mathbf{2},-1/2)$ & $(\mathbf{2},-1/2)$ & $(\mathbf{1},1)$\\ 
\multirow{2}{*}{$\quad\Gamma_4\quad \bigg\{$} & $\rho_i$ &
  $\rho_d$ & $\rho_u$ & $\rho_L$ & $\rho_{1,2,3}$ \\ 
& $k_i$ & $k_u$ & $k_d$ & $k_L$ & $k_{1,2,3}$ \\ 
\bottomrule
\end{tabular}
\caption{Transformation properties of the relevant MSSM chiral superfields under the gauge group and under the $\Gamma_4$ discrete modular symmetry. The $\rho_i$ denote the representations of the fields transforming under $S_4$ and the $k_i$ correspond to (minus) their modular weights.}
\label{tab:assignments}
\end{table}
%
We will keep the Higgs sector assignments trivial for simplicity.
We will also take lepton doublets (singlets) to transform as 
three (one) dimensional representations of $S_4$, as is customary.
Minimal models are then built by adhering to the following guidelines:
\begin{itemize}
\item No flavons are introduced,
\item Neutrino masses arise from the Weinberg operator,
\item Higgs multiplets transform trivially  $\rho_{u,d} \sim \mathbf{1}$
under $\Gamma_4$, with $k_{u,d} = 0$,
\item Lepton $\text{SU}(2)_L$ doublets transform as a triplet $\rho_L \sim \mathbf{3},\mathbf{3'}$ under $\Gamma_4$,
\item Lepton $\text{SU}(2)_L$ singlets transform as singlets $\rho_{1,2,3} \sim \mathbf{1},\mathbf{1'}$ under $\Gamma_4$, and
\item Lowest possible weights are chosen such that a rank 3 charged-lepton mass matrix $M_e$ is possible without imposing additional ``shaping'' symmetries.
\end{itemize}

Given the above conditions, we further expand the superpotential as: 
\begin{align}
W \,&=\,
\sum_i \alpha_i \, \big(E^c_i\, L\, H_d\, Y_\mathbf{2}^{a_i}\,Y_\mathbf{3'}^{b_i} \big)_\mathbf{1}
+ \frac{g}{\Lambda}\,\big(  L \,H_u\, L\, H_u\,
 Y_\mathbf{2}^c\,Y_\mathbf{3'}^d\big)_\mathbf{1}
 \,,
\label{eq:Wmin2}
\end{align}
%
where the integer, non-negative exponents of the modular 
forms satisfy $2(a_i+b_i) = k_L+k_i+k_d = k_L+k_i$
and $2(c+d) = k_L+k_u= k_L$.
In order to obtain some nontrivial structure in the Weinberg operator,
we assume that $k_L > 0$.
We explore in what follows two minimal choices of weights and representations,
corresponding to $k_L = 1$ (model I) and $k_L = 2$ (model II).

\subsection{Model I ($\boldsymbol{k_L=1}$)}

The choices $k_L = 1$ and $k_u = 0$
along with the fact that $\rho_L \sim \mathbf{3}$ or $\rho_L \sim \mathbf{3'}$
are enough to determine the structure of the Weinberg operator
and hence of the neutrino mass matrix $M_\nu$.

On the charged lepton side, the cancellation-of-weights
condition $2(a_i+b_i) = 1+k_i$ implies
that the $k_i$ are odd, $k_i \in \{-1,1,3,\ldots\}$.
To avoid having $\text{rank}(\lambda) < 3$,
one must make sure that different singlets can be extracted from 
each term in $W$.
If two lepton singlet superfields share the same weight and 
representation under $\Gamma_4$, the rank of the charged-lepton 
mass matrix is lower than its possible maximum.
If some $k_i = -1$, then $a_i=b_i=0$ and no singlet can be formed
for that generation.
If instead $k_i = 1$, a singlet can only be formed if 
$\rho_i \sim \mathbf{1'}$ ($\mathbf{1}$) when 
$\rho_L \sim \mathbf{3}$ ($\mathbf{3'}$).
We summarise in Table~\ref{tab:I} the weight and representation assignments of the minimal model allowing for $\text{rank}(\lambda) = 3$.

\begin{table}[h]
\centering
\renewcommand{\arraystretch}{1.2}
\begin{tabular}{lcccccccc}
\toprule
  & $H_u$ & $H_d$ && $L$ 
  & $E^c_1$ & $E^c_2$& $E^c_3$\\ 
\midrule 
 \multirow{2}{*}{$\rho_i\qquad$} & \multirow{2}{*}{$\mathbf{1}$} &
 \multirow{2}{*}{$\mathbf{1}$}
 && $\mathbf{3}$ & $\mathbf{1'}$  & $\mathbf{1}$  & $\mathbf{1'}$  \\
 & & && $\mathbf{3'}$ & $\mathbf{1}$  & $\mathbf{1'}$  & $\mathbf{1}$  \\[2mm]
 $k_i$    & $0$ & $0$ && $1$ & $1$ & $3$ & $3$ \\ 
\bottomrule
\end{tabular}
\caption{Transformation properties of chiral superfields under $\Gamma_4$, for the minimal model with $k_L = 1$ (model I). Both lines of lepton assignments lead to the same results.}
\label{tab:I}
\end{table}
%

With the weight assignments of Table~\ref{tab:I}, the superpotential reads:
\begin{equation}
\begin{aligned}
W^\text{I} &= 
  \alpha_1\, \big(E^c_1\,L\,Y_\mathbf{2}\big)_\mathbf{1}\,H_d
+ \alpha_2\, \big(E^c_1\,L\,Y_\mathbf{3'}\big)_\mathbf{1}\,H_d
\\
&\,
+ \beta_1\, \big(E^c_2\,L\,Y_\mathbf{2}^2\big)_\mathbf{1}\,H_d
+ \beta_2\, \big(E^c_2\,L\,Y_\mathbf{2}\,Y_\mathbf{3'}\big)_\mathbf{1}\,H_d
+ \beta_3\, \big(E^c_2\,L\,Y_\mathbf{3'}^2\big)_\mathbf{1}\,H_d
\\
&\,
+ \gamma_1\, \big(E^c_3\,L\,Y_\mathbf{2}^2\big)_\mathbf{1}\,H_d
+ \gamma_2\, \big(E^c_3\,L\,Y_\mathbf{2}\,Y_\mathbf{3'}\big)_\mathbf{1}\,H_d
+ \gamma_3\, \big(E^c_3\,L\,Y_\mathbf{3'}^2\big)_\mathbf{1}\,H_d
\\
&\,
+ \frac{g_1}{\Lambda}\,\big(L^2\,Y_\mathbf{2}\big)_\mathbf{1}\,H_u^2
+ \frac{g_2}{\Lambda}\,\big(L^2\,Y_\mathbf{3'}\big)_\mathbf{1}\,H_u^2
\,.
\label{eq:WI}
\end{aligned}
\end{equation}
%
Making use of the Clebsch-Gordan coefficients given in Appendix~\ref{app:cbc}
and of the $\Gamma_4\simeq S_4$ representation choices in Table~\ref{tab:I}
we find that only some terms in Eq.~\eqref{eq:WI} contain non-zero singlets.
We are left with
\begin{align}
W^\text{I}
&=
\alpha\, \big(E^c_1\,L\,Y_\mathbf{3'}\big)_\mathbf{1}\,H_d
+ \beta\, \big(E^c_2\,L\,Y_\mathbf{3'}^2\big)_\mathbf{1}\,H_d
+ \gamma\, \big(E^c_3\,L\,Y_\mathbf{2}\,Y_\mathbf{3'}\big)_\mathbf{1}\,H_d
+ \frac{g_1}{\Lambda}\,\big(L^2\,Y_\mathbf{2}\big)_\mathbf{1}\,H_u^2
\,,
\label{eq:WI2}
\end{align}
%
where $\alpha \equiv \alpha_2$, $\beta \equiv \beta_2+\beta_3$
and $\gamma \equiv \gamma_2$.
It is interesting to note that the constraints listed in 
Appendix~\ref{app:constraints} imply that the singlets
of the $\beta_2$ and $\beta_3$ terms coincide.
This superpotential gives rise to the mass matrices:
\begin{align}
M_\nu^\text{I} \,=\, \frac{2g_1 v_u^2}{\Lambda}\,
\begin{pmatrix}
0 & Y_2 & Y_1 \\
Y_2 & Y_1 & 0 \\
Y_1 & 0   & Y_2
\end{pmatrix}\,,
\label{eq:mnu1}
\end{align}
%
and
\begin{align}
M_e^\text{I} \,=\, v_d\,
\begin{pmatrix}
\alpha\,Y_3 & \alpha\,Y_5 & \alpha\,Y_4 \\ 
\beta\,(Y_1Y_4-Y_2Y_5) & \beta\,(Y_1Y_3-Y_2Y_4) & \beta\,(Y_1Y_5-Y_2Y_3) \\ 
\gamma\,(Y_1Y_4+Y_2Y_5) & \gamma\,(Y_1Y_3+Y_2Y_4) & \gamma\,(Y_1Y_5+Y_2Y_3) 
\end{pmatrix}^\dagger\,.
\label{eq:me}
\end{align}
%
Specifying values for the parameters 
$\alpha$, $\beta$, $\gamma$, and $\tau$
determines both mass matrices up to global factors
\footnote{
The parameters $\alpha$, $\beta$ and $\gamma$ can be made real 
through the rephasing of the singlet fields $E^c_i$. 
One of them may be taken outside of the matrix $M_e^\text{I}$ as a global factor.
It is assumed that the correct charged-lepton mass scale is reproduced
by an appropriate choice of this global factor, after $v_d$ has been specified.
},
and hence determines mass ratios as well as lepton mixing.

After performing a numerical search,
it seems this minimal model does not allow to reproduce known data.
As a benchmark, we find the point
$\alpha = 1$,
$\beta = 1.7\times 10^{-4}$,
$\gamma = 0.025$,
and $\tau = 0.19+0.99i$,
which accommodates a neutrino mass spectrum with inverted ordering,
but does not provide acceptable values for either $\sin^2 \theta_{12}$ or
$\sin^2 \theta_{13}$. For this point, we have
\begin{align}
\begin{array}{@{}r@{\,\,\simeq\,\,}l@{\qquad}r@{\,\,\simeq\,\,}l@{\qquad}r@{\,\,\simeq\,\,}l@{}}
\dfrac{m_e}{m_\mu}    & 0.0049\,,&
\sin^2 \theta_{12}    & 0.146\,,  &
\delta                & 1.23 \pi\,,
\\[6mm]
\dfrac{m_\mu}{m_\tau} & 0.0533\,,&
\sin^2 \theta_{13}    & 0.116\,, &
\alpha_{21}           & 1.92 \pi\,,
\\[6mm]
r                     & 0.0287\,,&
\sin^2 \theta_{23}    & 0.548\,, &
\alpha_{31}           & 0.50 \pi\,.
\end{array}
\end{align}
%
The obtained predictions are to be compared
with neutrino oscillation data and information on charged-lepton mass ratios,
which we collect in Table~\ref{tab:data}.

\begin{table}[t]
\centering
\renewcommand{\arraystretch}{1.2}
\begin{tabular}{lcc} 
\toprule
Parameter$\qquad\qquad$ & \multicolumn{2}{c}{Best-fit value and $1\sigma$ range} \\ 
\midrule
$m_e / m_\mu$ & \multicolumn{2}{c}{$0.0048 \pm 0.0002$} \\
$m_\mu / m_\tau$ & \multicolumn{2}{c}{$0.0565 \pm 0.0045$} \\ 
\midrule
& NO & IO \\
$\delta m^2/(10^{-5}\text{ eV})$ & \multicolumn{2}{c}{$7.34^{+0.17}_{-0.14}$} \\
$|\Delta m^2|/(10^{-3}\text{ eV})$ & $2.455^{+0.035}_{-0.032}$ & $2.441^{+0.033}_{-0.035}$ \\
$r \equiv \delta m^2/|\Delta m^2|$ & $0.0299\pm0.0008$ & $0.0301\pm0.0008$\\
$\sin^2\theta_{12}$ & $0.304^{+0.014}_{-0.013}$ & $0.303^{+0.014}_{-0.013}$ \\
$\sin^2\theta_{13}$ & $0.0214^{+0.0009}_{-0.0007}$ & $0.0218^{+0.0008}_{-0.0007}$ \\
$\sin^2\theta_{23}$ & $0.551^{+0.019}_{-0.070}$ & $0.557^{+0.017}_{-0.024}$ \\
$\delta/\pi$ & $1.32^{+0.23}_{-0.18}$  & $1.52^{+0.14}_{-0.15}$ \\
\bottomrule
\end{tabular}
\caption{Best-fit values and 1$\sigma$ ranges for 
neutrino oscillation parameters, obtained from the global analysis
of Ref.~\cite{Capozzi:2018ubv}, and for charged-lepton mass ratios,
given at the scale $2\times 10^{16}$ GeV with the $\tan \beta$ averaging
described in~\cite{Feruglio:2017spp}, obtained from Ref.~\cite{Ross:2007az}.
The parameters entering the definition of $r$ are $\delta m^2 \equiv m_2^2-m_1^2$
and $\Delta m^2 \equiv m_3^2 - (m_1^2+m_3^2)/2$.
The best-fit value and $1\sigma$ range of $\delta$
did not drive the numerical searches here reported.
}
\label{tab:data}
\end{table}

\subsection{Model II ($\boldsymbol{k_L=2}$)}

As before, the choices $k_L = 2$ and $k_u = 0$
along with the fact that $\rho_L \sim \mathbf{3}$ or $\rho_L \sim \mathbf{3'}$
are enough to determine the structure of $M_\nu$.

On the charged lepton side, the cancellation-of-weights
condition $2(a_i+b_i) = 2+k_i$ implies
that the $k_i$ are even, $k_i \in \{-2,0,2,\ldots\}$.
If some $k_i = -2$, then again $a_i=b_i=0$ and no singlet can be formed
for that generation.
If instead $k_i = 0$, a singlet can only be formed if $\rho_i \sim \mathbf{1'}$ ($\mathbf{1}$) when $\rho_L \sim \mathbf{3}$ ($\mathbf{3'}$).
We summarise in Table~\ref{tab:II} the weight and representation 
assignments of the minimal model allowing for $\text{rank}(\lambda) = 3$.

\begin{table}[h]
\centering
\renewcommand{\arraystretch}{1.2}
\begin{tabular}{lcccccccc}
\toprule
  & $H_u$ & $H_d$ && $L$ 
  & $E^c_1$ & $E^c_2$& $E^c_3$\\ 
\midrule 
 \multirow{2}{*}{$\rho_i\qquad$} & \multirow{2}{*}{$\mathbf{1}$} &
 \multirow{2}{*}{$\mathbf{1}$}
 && $\mathbf{3}$ & $\mathbf{1'}$  & $\mathbf{1}$  & $\mathbf{1'}$  \\
 & & && $\mathbf{3'}$ & $\mathbf{1}$  & $\mathbf{1'}$  & $\mathbf{1}$  \\[2mm]
 $k_i$    & $0$ & $0$ && $2$ & $0$ & $2$ & $2$ \\ 
\bottomrule
\end{tabular}
\caption{Transformation properties of chiral superfields under $\Gamma_4$, for the minimal model with $k_L = 2$ (model II). Both lines of lepton assignments lead to the same results.}
\label{tab:II}
\end{table}
%

With the weight assignments of Table~\ref{tab:II},
the charged-lepton Yukawa part of the superpotential
matches that of the case $k_L = 1$, leading to the same
charged-lepton mass matrix as the one given in Eq.~\eqref{eq:me},
i.e.~$M_e^\text{II} = M_e^\text{I}$.

The Weinberg operator part of the superpotential reads instead:
\begin{equation}
\begin{aligned}
W^\text{II} &\,\supset\,
 \frac{g_1}{\Lambda}\,\big(L^2\,Y_\mathbf{2}^2\big)_\mathbf{1}\,H_u^2
+ \frac{g_2}{\Lambda}\,\big(L^2\,Y_\mathbf{2}\,Y_\mathbf{3'}\big)_\mathbf{1}\,H_u^2
+ \frac{g_3}{\Lambda}\,\big(L^2\,Y_\mathbf{3'}^2\big)_\mathbf{1}\,H_u^2
\,.
\label{eq:WII}
\end{aligned}
\end{equation}
%
The first term in Eq.~\eqref{eq:WII} contributes
with two different non-zero singlets, since $L^2 \sim \mathbf{1} \oplus \mathbf{2} \oplus \mathbf{3}\oplus 0$, $Y_\mathbf{2}^2 \sim \mathbf{1}\oplus 0\oplus \mathbf{2}$, and
singlets may be obtained from both the $\mathbf{1}\otimes \mathbf{1}$
and the $\mathbf{2}\otimes \mathbf{2}$ combinations. 
The second term contributes with a third singlet,
as $Y_\mathbf{2}Y_\mathbf{3'} \sim \mathbf{3}\oplus \mathbf{3'}$
and thus a singlet is obtained from the $\mathbf{3}\otimes \mathbf{3}$
combination.
The third term contains all the three singlets: $\mathbf{1}\otimes \mathbf{1}$,
$\mathbf{2}\otimes \mathbf{2}$, and $\mathbf{3}\otimes \mathbf{3}$.
Due to the constraints in Appendix~\ref{app:constraints},
there are only three independent singlets which enter the Weinberg operator part of $W^\text{II}$. Explicitly, using the Clebsch-Gordan coefficients
of Appendix~\ref{app:cbc}:
\begin{equation}
\begin{aligned}
W^\text{II}
&\,\supset\,
\frac{1}{\Lambda}\,
\bigg[
g\,Y_1Y_2(L_1^2+2L_2L_3)
\\
&\qquad
+ g'\,\Big(Y_1^2(L_2^2+2L_1L_3)+Y_2^2(L_3^2 + 2L_2L_3)\Big)
\\
&\qquad
+ g''\,\Big(
(Y_1Y_4-Y_2Y_5)(L_1^2-L_2L_3)+
(Y_1Y_5-Y_2Y_3)(L_2^2-L_1L_3)
\\
&\qquad\qquad\qquad\qquad\qquad\qquad\qquad\qquad\,\,+
(Y_1Y_3-Y_2Y_4)(L_3^2-L_1L_2)
\Big)
\bigg]H_u^2
\,,
\label{eq:WII2}
\end{aligned}
\end{equation}
which gives rise to the mass matrix:
\begin{equation}
\begin{aligned}
M_\nu^\text{II} \,=\, \frac{2g' v_u^2}{\Lambda}&\,
\Bigg[\begin{pmatrix}
(g/g')Y_1Y_2 & Y_2^2 & Y_1^2 \\
Y_2^2 & Y_1^2 & (g/g')Y_1Y_2  \\
Y_1^2 & (g/g')Y_1Y_2    & Y_2^2
\end{pmatrix} \\
&+ \frac{1}{2}\frac{g''}{g'}
\begin{pmatrix}
2(Y_1Y_4-Y_2Y_5) & -(Y_1Y_3-Y_2Y_4) & -(Y_1Y_5-Y_2Y_3)\\
-(Y_1Y_3-Y_2Y_4) & 2(Y_1Y_5-Y_2Y_3) & -(Y_1Y_4-Y_2Y_5)\\
-(Y_1Y_5-Y_2Y_3) & -(Y_1Y_4-Y_2Y_5) & 2(Y_1Y_3-Y_2Y_4)
\end{pmatrix}
 \Bigg]
\,.
\label{eq:mnu2}
\end{aligned}
\end{equation}
%

Specifying values for the parameters 
$\alpha$, $\beta$, $\gamma$, $g/g'$, $g''/g'$ and $\tau$ 
determines both mass matrices up to global factors
\footnote{As before, 
$\alpha$, $\beta$ and $\gamma$ can be made real and one of them
may be taken outside of $M_e^\text{II}=M_e^\text{I}$ as a global factor.},
and hence determines mass ratios as well as lepton mixing.

Through numerical search,
we find that this minimal model is successful in accommodating the data.
We find a first benchmark, 
$\alpha = 0.12$,
$\beta = 4.2\times 10^{-4}$,
$\gamma = 1$,
$g/g' = 0.34 - 0.94 i$,
$g''/g' = 1.93 - 0.53 i$,
and $\tau = -0.1 + 1.2 i$,
which admits a neutrino mass spectrum with normal ordering, with
\begin{align}
\begin{array}{@{}r@{\,\,\simeq\,\,}l@{\qquad}r@{\,\,\simeq\,\,}l@{\qquad}r@{\,\,\simeq\,\,}l@{}}
\dfrac{m_e}{m_\mu}    & 0.0052\,,&
\sin^2 \theta_{12}    & 0.294\,,  &
\delta                & 0.32 \pi\,,
\\[6mm]
\dfrac{m_\mu}{m_\tau} & 0.0542\,,&
\sin^2 \theta_{13}    & 0.021\,, &
\alpha_{21}           & 0.96 \pi\,,
\\[6mm]
r                     & 0.0300\,,&
\sin^2 \theta_{23}    & 0.574\,, &
\alpha_{31}           & 1.59 \pi\,.
\end{array}
\end{align}
%
These results are in good agreement with the values of Table~\ref{tab:data},
except in what regards the experimental hint of $\delta \sim 3\pi/2$.

We find a second benchmark, 
$\alpha = 0.12$,
$\beta = 1$,
$\gamma = 4.5\times 10^{-4}$,
$g/g' = 15.1 + 7.7 i$,
$g''/g' = -0.40 - 1.08 i$,
and $\tau = 0.35 + 0.85 i$,
also admitting a neutrino mass spectrum with normal ordering,
for which the values of $\delta$ and of other measured parameters 
are less than $2\sigma$ away from the best-fit values
of Table~\ref{tab:data}:
\begin{align}
\begin{array}{@{}r@{\,\,\simeq\,\,}l@{\qquad}r@{\,\,\simeq\,\,}l@{\qquad}r@{\,\,\simeq\,\,}l@{}}
\dfrac{m_e}{m_\mu}    & 0.0045\,,&
\sin^2 \theta_{12}    & 0.278\,,  &
\delta                & 1.37 \pi\,,
\\[6mm]
\dfrac{m_\mu}{m_\tau} & 0.0557\,,&
\sin^2 \theta_{13}    & 0.021\,, &
\alpha_{21}           & 0.25 \pi\,,
\\[6mm]
r                     & 0.0296\,,&
\sin^2 \theta_{23}    & 0.480\,, &
\alpha_{31}           & 1.22 \pi\,.
\end{array}
\label{eq:2nd}
\end{align}
%
For this second benchmark, in order to fit the individual mass-squared 
differences $\delta m^2$ and $\Delta m^2$, we set the global 
factor $2g'v_u^2 / \Lambda \simeq 0.0024$ eV.
In this case, the neutrino masses read $m_1 \simeq 0.046$ eV,
$m_2 \simeq 0.047$ eV, and $m_3 \simeq 0.068 $ eV.
A distinctive feature of this framework is the prediction of
the Dirac and Majorana CPV phases. 
One is then in a position to extract a prediction for the 
effective Majorana mass $|\langle m\rangle|$ which controls the rate 
of neutrinoless double beta decay. Using the values in~\eqref{eq:2nd} we 
find $|\langle m\rangle|\simeq 0.043$ eV. This value can be probed by 
new-generation experiments which are working towards 
the $|\langle m \rangle| \sim 10^{-2}$ eV frontier.

In this setup, the correlations between pairs of mixing angles, phases and
mass ratios are non-trivial. The existence and the success of the above 
benchmarks warrant further exploration of the present framework.

\section{Summary and Conclusions}
\label{sec:conclusions}

In the present article we have explored 
the consequences of the presence of
modular invariance in the lepton sector 
via the action of the finite modular 
group $\Gamma_4\simeq S_4$
which is a subgroup of $\text{SL}(2,\amsmathbb{Z})$.

Fields carrying a non-trivial modular weight transform with 
a scale factor in addition to the usual unitary rotation.
To build an invariant theory under such transformations, 
special functions with the appropriate scaling properties  
need to be present in order to provide compensating factors. 
These functions are modular forms.
For a fixed scaling (modular) weight, they make up a 
finite-dimensional space. In section~\ref{sec:generators},
we have detailed the construction of a basis $\{Y_1,\ldots,Y_5\}$
for the lowest weight modular forms 
of level 4 (corresponding to $\Gamma_4\simeq S_4$), 
necessary to the generation of higher weight forms.
We have additionally shown how the weight 2 and 4 forms
organise themselves into different multiplets of $S_4$, and
that they satisfy non-trivial constraints which guarantee the correct
dimensionality of higher-weight spaces (see Appendix~\ref{app:constraints}).

We have then studied supersymmetric models of 
lepton masses and mixing
based on the breaking of the $\Gamma_4\simeq S_4$
modular invariance.
Focusing on minimality as a guiding principle,
we considered models where neutrino Majorana masses have their origin in
the Weinberg operator and where no flavons are introduced.
The expectation value of the modulus $\tau$
is the only source of symmetry breaking.

We study two minimal models, differing in the weight $-k_L$
of the lepton doublet under modular transformations.
While the first model (model I, $k_L = 1$) does not seem
to be able to reproduce the data, the second model (model II, $k_L = 2$)
successfully accommodates charged-lepton masses
and neutrino oscillation data. Deviations of individual parameters
from best-fit values are found to be below the $2\sigma$ level for
the benchmark of Eq.~\eqref{eq:2nd}.

Since in the present scheme
the mass matrices are fully determined by the VEV 
$\langle \tau \rangle$ and by superpotential parameters -- 
see Eqs.~\eqref{eq:mnu1},~\eqref{eq:me} and \eqref{eq:mnu2} --
the values of the Dirac and Majorana CPV phases can be predicted.
For the benchmark of Eq.~\eqref{eq:2nd},
which corresponds to a neutrino mass spectrum with normal ordering,
a prediction for the effective Majorana mass
$|\langle m\rangle|\simeq 0.043$ eV is possible.
This value can be probed by new-generation experiments
working towards the $|\langle m \rangle| \sim 10^{-2}$ eV frontier.

The predictions here derived are expected to hold at leading order.
There is a limited number of sources for deviations,
namely SUSY breaking and Kähler corrections.
The predictive power of the present approach 
and the existence of successful benchmarks
make this model building avenue worthy of future study.

\section*{Acknowledgements} 
%
We would like to thank M. Tanimoto for useful discussions.
This work was supported in part by the INFN
program on Theoretical Astroparticle Physics (TASP), 
by the European Union Horizon 2020 research and innovation programme
under the  Marie Sklodowska-Curie grants 674896 and 690575 (J.T.P.~and S.T.P.),
and by the World Premier International Research Center Initiative (WPI
Initiative), MEXT, Japan (S.T.P.).

\appendix
\numberwithin{equation}{section}
\section{$\mathbf{S_4}$ group theory}

\subsection{Presentation and basis}
\label{app:basis}

$S_4$ is the symmetric group of permutations of four objects. 
It contains $4! = 24$ elements and admits five irreducible representations 
$\mathbf{1}$, $\mathbf{1'}$, $\mathbf{2}$, $\mathbf{3}$ and $\mathbf{3'}$ 
(see, e.g.,~\cite{Ishimori:2010au}).
While a presentation in terms of three generators is common, 
it proves convenient
to consider here a presentation given in terms of two 
generators $S$ and $T$ (cf.~Eq.~\eqref{eq:present4}),
\begin{align}
S^2 \,=\, (ST)^3 \,=\, T^4 \,=\, \mathbb{1}\,.
\label{eq:present42}
\end{align}

We will use the group theoretical results of Ref.~\cite{Bazzocchi:2009pv}.
The two $S_4$ generators therein, which we denote here with primes,
satisfy $S'^4 = T'^3 = (S'T'^2)^2 = 1$. We define $S \equiv S'T'^2$, $T \equiv S'$,
which imply the inverse relations $S' = T$ and $T' = ST$.
Then, $S$ and $T$ furnish the presentation \eqref{eq:present42} of $S_4$,
useful to the discussion of modular invariance.
Making use of this identification and of the results in
Appendix A of Ref.~\cite{Bazzocchi:2009pv}, we
find an explicit basis for the irreducible representations of $S_4$,
which we employ in our discussion:
\begin{align}
\mathbf{1}:&\quad \rho(S)= 1 ,\quad \rho(T)= 1\,,\\
\mathbf{1'}:&\quad \rho(S)= -1,\quad \rho(T)= -1\,,\\
\mathbf{2}:&\quad \rho(S)= 
\begin{pmatrix}
0 & \omega \\ \omega^2 & 0
\end{pmatrix}
,\quad \rho(T)= 
\begin{pmatrix}
0 & 1 \\ 1 & 0
\end{pmatrix}
\,,\\
\mathbf{3}:&\quad \rho(S)=
\frac{1}{3}
\begin{pmatrix}
-1 & 2 \omega^2 & 2 \omega \\
 2 \omega & 2  & -\omega^2  \\
2 \omega^2 & -\omega & 2  
\end{pmatrix} 
,\quad \rho(T)=
\frac{1}{3}
\begin{pmatrix}
-1 & 2 \omega & 2 \omega^2 \\
 2 \omega & 2 \omega^2 & -1  \\
2 \omega^2 & -1 & 2 \omega  
\end{pmatrix}  \,,\\
\mathbf{3'}:&\quad \rho(S)= 
-\frac{1}{3}
\begin{pmatrix}
-1 & 2 \omega^2 & 2 \omega \\
 2 \omega & 2  & -\omega^2  \\
2 \omega^2 & -\omega & 2  
\end{pmatrix} 
,\quad \rho(T)=
-\frac{1}{3}
\begin{pmatrix}
-1 & 2 \omega & 2 \omega^2 \\
 2 \omega & 2 \omega^2 & -1  \\
2 \omega^2 & -1 & 2 \omega  
\end{pmatrix}\,,
\end{align}
where as usual $\omega = e^{2\pi i/3}$.

\subsection{Clebsch-Gordan Coefficients}
\label{app:cbc}
After establishing a dictionary between presentations (see previous subsection),
we can directly use the Clebsch-Gordan coefficients from Ref.~\cite{Bazzocchi:2009pv},
since no change of basis on the representation matrices has been performed. 
We reproduce the coefficients here for completeness. Entries of each multiplet entering the tensor product are denoted by $\alpha_i$ and $\beta_i$.
\begin{align}
\begin{array}{@{}c@{{}\,\otimes\,{}}c@{{}\,\,=\,\,{}}l@{\quad\sim\quad}l@{}}
\mathbf{1}&\mathbf{r}&\mathbf{r} &\alpha\,\beta_i \\[2mm]
\mathbf{1'}&\mathbf{1'}&\mathbf{1}&\alpha\,\beta \\[2mm]
\mathbf{1'}&\mathbf{2}&\mathbf{2}&
\begin{pmatrix}
\alpha\,\beta_1\\
-\alpha\,\beta_2 
\end{pmatrix}\\[4mm]
\mathbf{1'}&\mathbf{3}&\mathbf{3'}&
\begin{pmatrix}
\alpha\,\beta_1 \\
\alpha\,\beta_2 \\
\alpha\,\beta_3
\end{pmatrix} \\[6mm]
\mathbf{1'}&\mathbf{3'}&\mathbf{3}&
\begin{pmatrix}
\alpha\,\beta_1 \\
\alpha\,\beta_2 \\
\alpha\,\beta_3
\end{pmatrix} 
\end{array}
\end{align}

\begin{align}
\begin{array}{@{}c@{{}\,\otimes\,{}}c@{{}\,\,=\,\,{}}ll@{}}
\mathbf{2}&\mathbf{2}&\mathbf{1}\,\oplus\, \mathbf{1'}\,\oplus\, \mathbf{2} &
\left\{\begin{array}{@{}l@{\quad\sim\quad}l@{}}
\quad \mathbf{1}  & \alpha_1\beta_2+\alpha_2\beta_1\\[2mm]
\quad \mathbf{1'} & \alpha_1\beta_2-\alpha_2\beta_1\\[2mm]
\quad \mathbf{2}  & \begin{pmatrix} \alpha_2\,\beta_2\\
                                    \alpha_1\,\beta_1 
                    \end{pmatrix}
\end{array}\right.
\\[13mm]
\mathbf{2}&\mathbf{3}&\mathbf{3}\,\oplus\,\mathbf{3'}&
\left\{\begin{array}{@{}l@{\quad\sim\quad}l@{}}
\quad \mathbf{3}   & \begin{pmatrix} \alpha_1\,\beta_2 + \alpha_2\,\beta_3 \\
                                     \alpha_1\,\beta_3 + \alpha_2\,\beta_1 \\
                                     \alpha_1\,\beta_1 + \alpha_2\,\beta_2
                     \end{pmatrix}\\[6mm]
\quad \mathbf{3'}  & \begin{pmatrix} \alpha_1\,\beta_2 - \alpha_2\,\beta_3 \\
                                     \alpha_1\,\beta_3 - \alpha_2\,\beta_1 \\
                                     \alpha_1\,\beta_1 - \alpha_2\,\beta_2
                     \end{pmatrix}
\end{array}\right.
\\[16mm]
\mathbf{2}&\mathbf{3'}&\mathbf{3}\,\oplus\,\mathbf{3'}&
\left\{\begin{array}{@{}l@{\quad\sim\quad}l@{}}
\quad \mathbf{3}   & \begin{pmatrix} \alpha_1\,\beta_2 - \alpha_2\,\beta_3 \\
                                     \alpha_1\,\beta_3 - \alpha_2\,\beta_1 \\
                                     \alpha_1\,\beta_1 - \alpha_2\,\beta_2
                     \end{pmatrix}\\[6mm]
\quad \mathbf{3'}  & \begin{pmatrix} \alpha_1\,\beta_2 + \alpha_2\,\beta_3 \\
                                     \alpha_1\,\beta_3 + \alpha_2\,\beta_1 \\
                                     \alpha_1\,\beta_1 + \alpha_2\,\beta_2
                     \end{pmatrix}
\end{array}\right.
\end{array}
\end{align}

\begin{align}
\begin{array}{@{}cll@{}}
\mathbf{3}\,\otimes\,\mathbf{3}\,=\,\mathbf{3'}\,\otimes\,\mathbf{3'}\,=\,\mathbf{1}\,\oplus\, \mathbf{2}\,\oplus\, \mathbf{3}\,\oplus\, \mathbf{3'} &
\left\{\begin{array}{@{}l@{\quad\sim\quad}l@{}}
\quad \mathbf{1}  & \alpha_1\beta_1+\alpha_2\beta_3+\alpha_3\beta_2\\[2mm]
\quad \mathbf{2}  & \begin{pmatrix}
                      \alpha_2\beta_2+\alpha_1\beta_3+\alpha_3\beta_1 \\
                      \alpha_3\beta_3+\alpha_1\beta_2+\alpha_2\beta_1
                    \end{pmatrix}\\[4mm]
\quad \mathbf{3}   & \begin{pmatrix} 
                       2\alpha_1\beta_1-\alpha_2\beta_3-\alpha_3\beta_2 \\
                       2\alpha_3\beta_3-\alpha_1\beta_2-\alpha_2\beta_1 \\
                       2\alpha_2\beta_2-\alpha_1\beta_3-\alpha_3\beta_1
                     \end{pmatrix}\\[6mm]
\quad \mathbf{3'}  & \begin{pmatrix}
                       \alpha_2\beta_3-\alpha_3\beta_2 \\
                       \alpha_1\beta_2-\alpha_2\beta_1 \\
                       \alpha_3\beta_1-\alpha_1\beta_3
                     \end{pmatrix}
\end{array}\right.
\end{array}
\end{align}

\begin{align}
\begin{array}{@{}cll@{}}
\mathbf{3}\,\otimes\,\mathbf{3'}\,=\,\mathbf{1'}\,\oplus\, \mathbf{2}\,\oplus\, \mathbf{3}\,\oplus\, \mathbf{3'} &
\left\{\begin{array}{@{}l@{\quad\sim\quad}l@{}}
\quad \mathbf{1'} & \alpha_1\beta_1+\alpha_2\beta_3+\alpha_3\beta_2\\[2mm]
\quad \mathbf{2}  & \begin{pmatrix}
                      \alpha_2\beta_2+\alpha_1\beta_3+\alpha_3\beta_1 \\
                      -\alpha_3\beta_3-\alpha_1\beta_2-\alpha_2\beta_1
                    \end{pmatrix}\\[4mm]
\quad \mathbf{3}   & \begin{pmatrix} 
                       \alpha_2\beta_3-\alpha_3\beta_2 \\
                       \alpha_1\beta_2-\alpha_2\beta_1 \\
                       \alpha_3\beta_1-\alpha_1\beta_3
                     \end{pmatrix}\\[6mm]
\quad \mathbf{3'}  & \begin{pmatrix}
                       2\alpha_1\beta_1-\alpha_2\beta_3-\alpha_3\beta_2 \\
                       2\alpha_3\beta_3-\alpha_1\beta_2-\alpha_2\beta_1 \\
                       2\alpha_2\beta_2-\alpha_1\beta_3-\alpha_3\beta_1
                     \end{pmatrix}
\end{array}\right.
\end{array}
\end{align}

\section{$\boldsymbol{q}$-expansions of lowest weight $\boldsymbol{N=4}$ modular forms}
\label{app:qexp}

The five linearly independent modular forms in Eq.~\eqref{eq:Yi} admit the expansions:
\begin{align}
-\frac{8i}{3\pi}\,& Y_1(\tau) \,=\,
1-24y-72y^2+288y^3+216y^4+\ldots
\,,\\[2mm]
-\frac{8i}{3\pi}\,& Y_2(\tau) \,=\,
1+24y-72y^2-288y^3+216y^4+\ldots
\,,\\[2mm]
\frac{4i}{\pi}\,& Y_3(\tau) \,=\,
1 - 8 z + 64 z^3 + 32 z^4 + 192 z^5 - 512 z^7 + 384 z^8+\ldots
\,,\\[2mm]
\frac{2i}{\pi}\,& \left[Y_4(\tau)+Y_5(\tau)\right] \,=\,
1 + 4 z - 32 z^3 + 32 z^4 - 96 z^5 + 256 z^7 + 384 z^8 +\ldots
\,,\\[2mm]
\frac{i}{\pi}\,& \left[Y_4(\tau)-Y_5(\tau)\right] \,=\,
2\sqrt{3}\, z\left( 1 + 8 z^2 - 24 z^4 - 64 z^6 +\ldots \right )
\,,
\end{align}
where $y \equiv i \sqrt{q/3}$, $z \equiv e^{i\pi/4}(q/4)^{1/4}$, and as usual $q = e^{2\pi i\,\tau}$.

\section{Forms of higher weight and constraints}
\label{app:constraints}
Through tensor products of $Y_\mathbf{2}$ and $Y_\mathbf{3'}$,
one can find the multiplets:
\begin{equation}
\begin{aligned}
Y^{(4)}_\mathbf{1}    &\,=\, Y_1 Y_2 \,\sim\, \mathbf{1}  \,,\\[2mm]
{Y^{(4)}_\mathbf{1}}' &\,=\, Y_3^2 + 2\,Y_4 Y_5 \,\sim\, \mathbf{1}  \,,\\[2mm]
{Y^{(4)}_\mathbf{2}}  &\,=\, (Y_2^2,\, Y_1^2)^T \,\sim\, \mathbf{2}  \,,\\[2mm]
{Y^{(4)}_\mathbf{2}}' &\,=\, (Y_4^2+2Y_3Y_5,\, Y_5^2+2Y_3Y_4)^T \,\sim\, \mathbf{2}  \,,\\[2mm]
{Y^{(4)}_\mathbf{3}}  &\,=\, (Y_1Y_4-Y_2Y_5,\,Y_1Y_5-Y_2Y_3,\,Y_1Y_3-Y_2Y_4)^T \,\sim\, \mathbf{3}  \,,\\[2mm]
{Y^{(4)}_\mathbf{3}}'  &\,=\, (Y_3^2-Y_4Y_5,\,Y_5^2-Y_3Y_4,\,Y_4^2-Y_3Y_5)^T \,\sim\, \mathbf{3}  \,,\\[2mm]
{Y^{(4)}_\mathbf{3'}}  &\,=\, (Y_1Y_4+Y_2Y_5,\,Y_1Y_5+Y_2Y_3,\,Y_1Y_3+Y_2Y_4)^T \,\sim\, \mathbf{3'}
\,.
\label{eq:15}
\end{aligned}
\end{equation}
Not all of these multiplets are expected to be independent.
Indeed, from the $q$-expansions of the $Y_i(\tau)$ given 
in Appendix~\ref{app:qexp} we find 6 constraints between the 15 different $Y_i(\tau)Y_j(\tau)$ products:
\begin{align}
\begin{array}{r@{\,\,=\,\,}l@{\qquad}r@{\,\,=\,\,}l}
\dfrac{1}{3}\left(Y_3^2 + 2 Y_4 Y_5\right) & Y_1 Y_2\,,
&
-\dfrac{1}{\sqrt{3}}\left(Y_3^2 - Y_4 Y_5\right) & Y_1 Y_4 - Y_2 Y_5 \,,\\[3mm]
\dfrac{1}{3}\left(Y_4^2 + 2 Y_3 Y_5\right) & Y_2^2 \,,
&
-\dfrac{1}{\sqrt{3}}\left(Y_5^2 - Y_3 Y_4\right) & Y_1 Y_5 - Y_2 Y_3 \,,\\[3mm]
\dfrac{1}{3}\left(Y_5^2 + 2 Y_3 Y_4\right) & Y_1^2 \,,
&
-\dfrac{1}{\sqrt{3}}\left(Y_4^2 - Y_3 Y_5\right) & Y_1 Y_3 - Y_2 Y_4 \,.
\end{array}
\label{eq:constraints}
\end{align}
These constraints imply that
$Y^{(4)}_\mathbf{1}$ and ${Y^{(4)}_\mathbf{1}}'$,
$Y^{(4)}_\mathbf{2}$ and ${Y^{(4)}_\mathbf{2}}'$,
and $Y^{(4)}_\mathbf{3}$ and ${Y^{(4)}_\mathbf{3}}'$
 in Eq.~\eqref{eq:15} denote the same multiplets, and only one of each pair is kept
in our discussion, cf.~Eq.~\eqref{eq:9}.

In Ref.~\cite{Feruglio:2017spp} it is argued that 
the presence of a covariant constraint similar to
the ones given in Eq.~\eqref{eq:constraints}
signals the non-linear realisation of the discrete symmetry.

\bibliographystyle{JHEPwithnote}
\bibliography{bibliography}

\end{document}